\documentstyle[pre,aps,twocolumn,epsf]{revtex}
\begin{document}
\title{Analytic Theory of fractal growth patterns in 2 dimensions}
\author{Benny Davidovitch and Itamar Procaccia}
\address{Dept. of Chemical Physics, The Weizmann Institute of
Science, Rehovot 76100, Israel}

\date{version of \today}
\maketitle

{\bf Diffusion Limited Aggregation (DLA) is a model of fractal growth
that was introduced in 1981 by Witten and Sander \cite{81WS}. It had since
attained a paradigmatic status
due to its simplicity and its underlying role for a variety of pattern
forming processes including dielectric breakdown \cite{84NPW}, viscous
fingering\cite{85NDS},
electro-chemical deposition \cite{86GBCS,89BMT}, dendritic and snowflake
growth \cite{86KV}, chemical
dissolution \cite{87DL}, geological phenomena \cite{89FSD} and certain
biological
phenomena such as bacterial growth \cite{94BSTCCV} and viscous fingering
through
gastric mucin \cite{92Mucin}. The algorithm begins with
fixing one particle at the center of coordinates in $d$-dimensions, and follows
the creation of a cluster by releasing random walkers from infinity,
allowing them to walk around until they hit any particle belonging to the
cluster. The fundamental difficulty of such growth processes is that their
mathematical
description calls for solving equations with boundary conditions on a
complex, evolving interface. Despite tremendous efforts this and related
difficulties defied all attempts to develop analytic theory
of DLA. In fact, even the numerical estimates \cite{83Mea}
of the fractal dimension $D$ of DLA turned out to converge extremely
slowly with the number of particles $n$ of the cluster, leading even to
speculations \cite{Mandel} that
asymptotically the clusters were plane filling (i.e. $D=2$).
In this Letter we offer a theory for fractal growth patterns in 2-d,
including DLA as a particular case. In this theory the fractal dimension of
the asymptotic cluster manifests itself as a renormalization exponent
observable
already at very early growth stages. Using early stage dynamics we compute
$1.6896<D<1.7135$,
and explain why traditional numerical estimates converged so slowly. The
present theory
is equally applicable to other fractal growth processes in 2-dimensions,
and we discuss similar
computations for such growth models as well}.

For {\em continuous time} processes in 2-dimensions the above mentioned
difficulty was efficiently
dealt with in the past \cite{45Pol,58ST,84SB} by considering the conformal
map from the exterior of the
unit circle in the complex plane to the exterior of the (simply connected)
growing pattern. In
this way the ``interface" in the mathematical plane remains forever simple, and
the complexity of the evolving interface is delegated to
the dynamics of the conformal map. For {\em discrete particle growth} such
a language
was developed recently  \cite{98HL,99DHOPSS,99DP,00DFHP},
showing that a large variety of fractal clusters in two dimensions can be
grown
by iterating conformal
maps. In this Letter we employ this language to show that the fractal dimension
of the asymptotically large clusters has a surprising and useful role as a
renormalization
exponent in a rescaling theory of these clusters in their early growth phases.
This finding allows us to compute the fractal dimension to desired accuracy.

Once a fractal object is well developed, it is extremely difficult to find
a conformal map from a smooth region to its boundary, simply because the
conformal
map is terribly singular on the tips of a fractal shape. The derivative of
the inverse map is the
growth probability for a random walker to hit the interface (known as
the ``harmonic measure") which has been shown to be a multifractal measure
\cite{86HMP}
characterized by infinitely many exponents \cite{83HP,86HJKPS}.
Accordingly, in the present
approach one grows the cluster by iteratively constructing the conformal
map starting from
a smooth initial interface. Consider
$\Phi^{(n)}(w)$ which conformally maps the exterior of the unit circle
$e^{i\theta}$ in the
mathematical $w$--plane onto the complement of the (simply-connected)
cluster of $n$ particles in the physical $z$--plane
\cite{98HL,99DHOPSS,99DP,00DFHP}.
The unit circle is
mapped to the boundary of the cluster. The map $\Phi^{(n)}(w)$ is
made from compositions of elementary maps $\phi_{\lambda,\theta}$,
\begin{equation}
\Phi^{(n)}(w) = \Phi^{(n-1)}(\phi_{\lambda_{n},\theta_{n}}(w)) \ ,
\label{recurs}
\end{equation}
where the elementary map $\phi_{\lambda,\theta}$ transforms the unit
circle to a circle with a ``bump" of linear size $\sqrt{\lambda}$ around
the point $w=e^{i\theta}$. An example of a good elementary map
$\phi_{\lambda,\theta}$ was
proposed in \cite{98HL}, endowed with a parameter
$a$ in the range $0< a < 1$, determining the shape of the bump. We employ
$a=2/3$ which is consistent with
semicircular bumps. Accordingly the map $\Phi^{(n)}(w)$ adds on a
new bump to the image of the unit circle under $\Phi^{(n-1)}(w)$. The
bumps in the $z$-plane simulate the accreted particles in
the physical space formulation of the growth process. Since we want
to have {\em fixed size} bumps in the physical space, say
of fixed area $\lambda_0$, we choose in the $n$th step
\begin{equation}
\lambda_{n} = \frac{\lambda_0}{|{\Phi^{(n-1)}}' (e^{i \theta_n})|^2} \ .
\label{lambdan}
\end{equation}
The recursive dynamics can be represented as iterations
of the map $\phi_{\lambda_{n},\theta_{n}}(w)$,
\begin{equation}
\Phi^{(n)}(w) =
\phi_{\lambda_1,\theta_{1}}\circ\phi_{\lambda_2,\theta_{2}}\circ\dots\circ
\phi_{\lambda_n,\theta_{n}}(\omega)\ . \label{comp}
\end{equation}

\begin{figure}
\epsfxsize=7.0truecm
\epsfysize=7.0truecm
\epsfbox{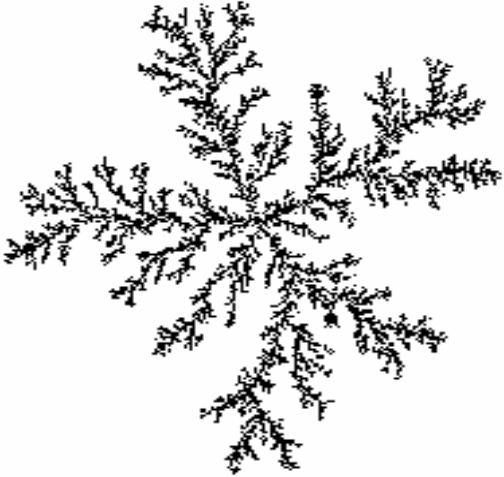}
\caption{A DLA cluster, $n=100 000$.}
\label{DLAcluster}
\end{figure}

The difference between various growth models will manifest itself in the
different itineraries $\{\theta_1\cdots \theta_n\}$.
To grow a DLA we have to choose random positions $\theta_n$.
This way we accrete fixed size bumps in the physical plane according
to the harmonic measure (which is transformed into a uniform
measure by the analytic inverse of $\Phi^{(n)}$).
The DLA cluster is fully determined by the stochastic itinerary
$\{\theta_k\}_{k=1}^n$. In Fig.~\ref{DLAcluster} we present a typical DLA
cluster
grown by this method to size $n=$100 000.
\narrowtext
\narrowtext
\begin{figure}
\epsfxsize=7.0truecm
\epsfysize=7.0truecm
\epsfbox{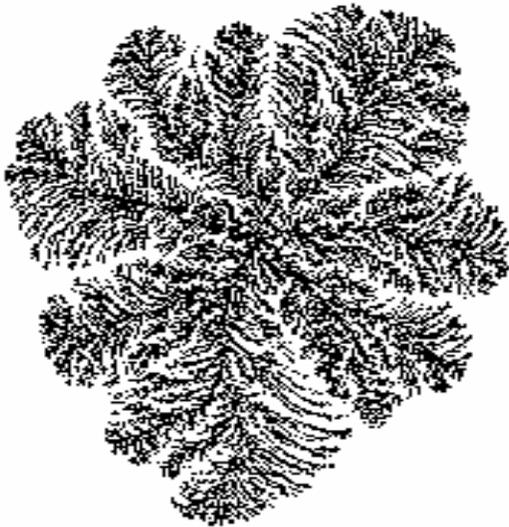}
\caption{A deterministic cluster with a Golden Mean itinerary, $n=100 000$.}
\label{gmcluster}
\end{figure}
Other fractal clusters can be obtained by choosing a non-random
itinerary \cite{00DFHP}. A beautiful family of growth patterns is obtained from
quasi-periodic itineraries,
\begin{equation}
\theta_{k+1}=\theta_k+2\pi W \ , \label{quasip}
\end{equation}
where $W$ is a quadratic irrational number. An example is shown in
Fig.~\ref{gmcluster}, in which $W$ is the golden mean $(\sqrt{5}+1)/2$.
In \cite{00DFHP} it was argued that itineraries obtained from
(\ref{quasip}) using for $W$ other values of quadratic irrationals lead to
clusters of different appearance but the same dimension, which
was estimated numerically to be $D=1.86\pm 0.03$. In the same
paper other deterministic itineraries (not obtained from circle maps)
where shown to lead to clusters with different dimensions. One (trivial)
example
that is nevertheless useful for our consideration below is the itinerary
$\theta_k=0$ for all $k$. Such an itinerary grows a 1 dimensional
wire of width $\sqrt{\lambda_0}$.

The great advantage of the availability of a conformal map
is that it affords us analytic power that is not obtainable
otherwise. To understand this consider the Laurent expansion of
$\Phi^{(n)}(w)$ :
\begin{equation}
\Phi^{(n)}(w) = F^{(n)}_{1} w + F^{(n)}_{0} + F^{(n)}_{-1}w^{-1} +
F^{(n)}_{-2}w^{-2} + \dots
\label{eq-laurent-f}
\end{equation}
The recursion equations for the Laurent coefficients of $\Phi^{(n)}(w)$
can be obtained analytically, and in particular one shows that \cite{98HL}
\begin{equation}
F_1^{(n)}  = \prod_{k=1}^{n}  [1+\lambda_k]^a  \,. \label{F1a}
\end{equation}
The first Laurent coefficient $F_1^{(n)}$ has a distinguished role in
determining
the fractal dimension of the cluster, being identical to the
Laplace radius which is the radius of a charged disk having the
same field far away as the charged cluster \cite{99DHOPSS}. Moreover,
defining $R_n$ as the minimal radius of
all circles in $z$ that contain the $n$-cluster, one can prove that
\cite{83Dur}
\begin{equation}
R_n\le 4F_1^{(n)} \ . \label{RvsF}
\end{equation}
Accordingly one expects that for sufficiently large clusters (to be made
precise below)
\begin{equation}
F^{(n)}_1\sim n^{1/D}\sqrt{\lambda_0} \ , \quad n\to \infty \ ,
\label{eq-scalingrad}
\end{equation}
as $\sqrt{\lambda_0}$ remains the only length scale in the problem when the
radius of the cluster is much larger than the radius of the initial
smooth interface (which we take as the
unit circle in this discussion, $\Phi^{(0)}(\omega)=\omega)$).

These observations lead now to the central development of this Letter,
and to the most important result. Consider a renormalization
process in which we fix the initial smooth interface, but change
$\lambda_0$, and then rescale $n$ such
as to get the ``same" cluster. Of course we need to specify what do
we mean by the ``same" cluster, and a natural requirement is that
the electrostatic field on coarse scales (i.e. far from the cluster)
will remain {\em invariant}.
In other words, we should require the
invariance of the Laplace radius $F_1^{(n)}$ (and possibly
of additional low order Laurent coefficients) under renormalization.
Clearly, for a given itinerary $\{\theta_k\}_{k=1}^n$, $F_1^{(n)}$ is
a function of $n$ and $\lambda_0$ only. Accordingly, considering
Eq.(\ref{eq-scalingrad}), we note that such a renormalization
process can reach a fixed point if and only if $F_1^{(n)}(\lambda_0)$ attains
a nontrivial fixed point function $F_1^*$ of the single ``scaling" variable
$x=\sqrt{\lambda_0}n^{1/D}$. Obviously in the asymptotic limit $x \gg 1$
$F_1^{(n)}(\lambda_0)$ must converge to $F_1^*$ which is {\em linear} in
$x$ in this regime.
The main new findings of this Letter are that $F_1^*$ exists as a nonlinear
function of $x$,
and that $F_1^{(n)}(\lambda_0)$ converges
(within every universality class) to its fixed point function $F_1^*$
already for $x\ll 1$.

\begin{figure}
\hskip +0.5 cm
\epsfxsize=7.0truecm
\epsfysize=7.0truecm
\epsfbox{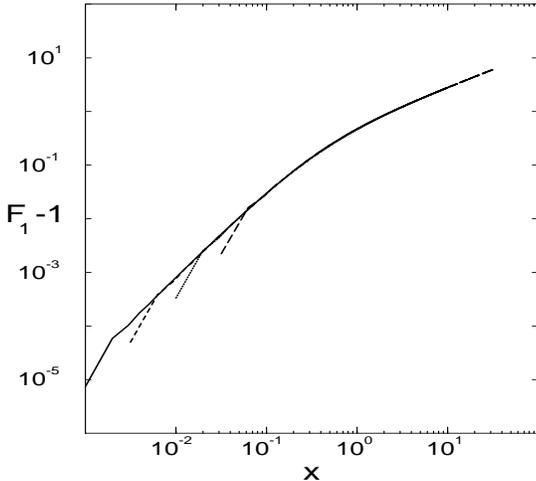}
\caption{The one dimensional wire. We plot $F_1(x)-1$ vs.
$x=n{\lambda^{1/2}_0}$ and demonstrate the
convergence to the asymptotic nonlinear function $F_1^{*}$. The values of
$\lambda_0$ used
are $10^{-3}$, $10^{-4}$, $10^{-5}$ and $10^{-6}$.}
\label{wirecollapse}
\end{figure}

In principle one can demonstrate the convergence to $F_1^*$ analytically.
This is easy to do in the case of the degenerate itinerary growing a wire.
In this case we can demonstrate convergence after the addition of
2-3 bumps, even in the limit $\lambda_0\to 0$, see Fig.~\ref{wirecollapse}
. For $x\to 0$
$F_1^*(x)-1\sim x^2$ in this case.
\narrowtext

\begin{figure}
\hskip +0.5 cm
\epsfxsize=7.0truecm
\epsfysize=7.0truecm
\epsfbox{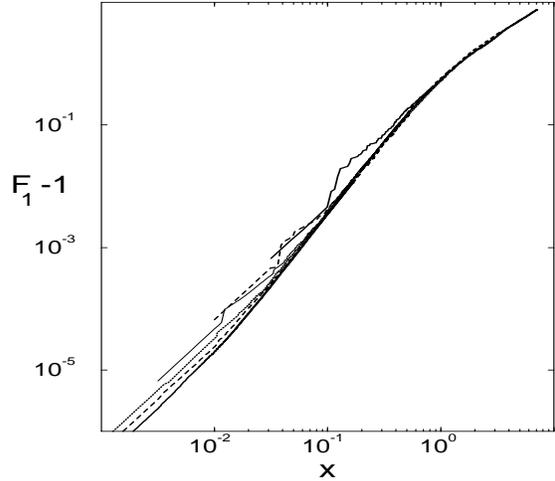}
\caption{Convergence to $F_1^*$ for the DLA. We plot
$F_1(x)-1$ vs. $x=n^{1/D}\lambda^{1/2}_0$ with $D=1.7$. The values of
$\lambda_0$ used
are $10^{-3}$ (upper solid line), $10^{-4}$ (upper dashed), $10^{-5}$ (thin
solid)
$10^{-6}$ (dotted), $10^{-7}$ (lower dashed) and $10^{-8}$ (lower solid).
The fixed point
function $F_1^*$ is best approximated by the lowest curve in the figure.
Convergence to the asymptotic nonlinear function $F_1^*$ is seen
for smaller $x$ values when $\lambda_0$ decreases. }
\label{DLAcollapse}
\end{figure}

\begin{figure}
\hskip +0.5 cm
\epsfxsize=7.0truecm
\epsfysize=7.0truecm
\epsfbox{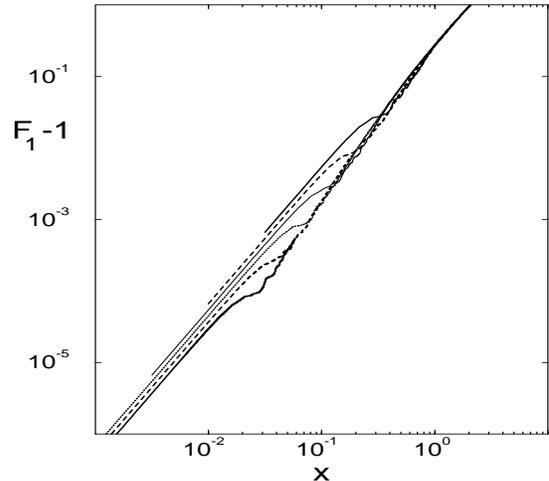}
\caption{Convergence to $F_1^*$ for the Golden Mean itinerary. Data are same
as in Fig. 4 but $x=n^{1/D}\lambda^{1/2}_0$ with $D=1.83$. The fixed point
function $F_1^*$ is best approximated by the lowest curve for
$x>3\times10^{-2}$
where the data with $\lambda_0=10^{-8}$ converges to $F_1^*$.}
\label{GMcollapse}
\end{figure}

For other nontrivial itineraries it becomes increasingly cumbersome to
demonstrate the
convergence by hand. With the assistance of the machine we can
demonstrate the convergence in all the other cases. In
Fig.~\ref{DLAcollapse} we present
$F_1^{(n)}(\lambda_0)-1$ as a function of $x$ for a typical DLA itinerary
and for values of $\lambda_0$ ranging between $10^{-8}$ to $10^{-3}$. We note
that for $\lambda_0\to 0$ the convergence to the fixed point function
is obtained infinitesimally close to the initial circle for which  $F_1^{(n=0)}=1$.
In fact, data collapse (with $D$ chosen right) for this itinerary, as
well as for all other nontrivial itineraries,
is obtained for $n\ge n_c$ where $n_c\approx 2\pi/\sqrt{\lambda_0}$. This
is the number of bumps
required to obtain one-layer coverage of the original circular interface.
Obviously
$n^{1/D}_c \sqrt{\lambda_0}\to 0$ for $\lambda_0\to 0$, demonstrating the
convergence to $F_1^*$ for $x\ll 1$. In Fig.\ref{GMcollapse} we exhibit the
convergence for the Golden Mean itinerary. Note that the fixed point
functions are different, and they both differ from the wire
case. The main point of this analysis is that convergence to $F_1^*$ can be
obtained for $x$ arbitrarily small by going to the limit $\lambda_0\to 0$.
\narrowtext
\narrowtext
The existence of a fixed point function translates immediately to
a calculational scheme. Consider a given itinerary $\{\theta_k\}_{k=1}^N$ of
one of the above classes, and calculate $F_1^{(n)}(\lambda_0)$ for
$N>n>n_c(\lambda_0)$.
Rescale now $\lambda_0\to \lambda_0/s$, and calculate $F_1^{(n')}(\lambda_0/s)$
for $N>n'>n_c(\lambda_0/s)$. We can compute $D$ from finding the value $n'$
which preserves the Laplace radius
under rescaling of $\lambda_0$ by $s$:
\begin{equation}
\left(\frac{n'}{n}\right)^{1/D} =\sqrt{s} \ . \label{zehu}
\end{equation}
Since $F_1^*$ is monotonic (as is immediately seen from Eq.(\ref{F1a})), there
is only one solution to this equation,
\begin{equation}
D = \frac{2[\log{n'}-\log{n}]}{\log{s}} \ . \label{D}
\end{equation}

As a first example consider the wire case. Computing $F_1^{(10)}(10^{-6})$ we
find that for $\lambda_0=10^{-5}$ the same value of $F_1^{(n)}$ is obtained
for $n$ between 3 and 4. Eq.(\ref{D}) with $s=10$ then predicts
$0.796<D<1.045$. Repeating
for $F_1^{(100)}(10^{-6})$ we find the same value of $F_1^{(n)}(10^{-5})$ for
$n$ between 31 and 32. From Eq.(\ref{D}) $0.9897<D<1.0173$. We stress that this
precision is obtained when $F_1^{(n)}-1$ is still 0.0133! Lastly, using
$F_1^{(500)}(10^{-6})=1.18254$ we compute $0.9951<D<1.0006$, and any desired
accuracy can be achieved by increasing $n$. The reader should note that the
fractal
dimension of the asymptotic cluster ($D=1$ in this case) can be extracted
from growth
events infinitesimally close to the initial unit circle by decreasing
$\lambda_0$ (even without
increasing $n$ in this specific case). Secondly we consider the
deterministic itinerary
(\ref{quasip}) with Golden Mean winding number. Using values of
$F_1^{(n)}(10^{-6})$
and $F_1^{(n)}(10^{-5})$ between 1.10 and 1.20 we can bound the dimension
of the
cluster to $1.8305<D<1.8380$. Note that in this case we need to have at least
one layer covering which is obtained only when $n_c\approx 2\pi/\sqrt{\lambda_0}$.
Finally, we consider DLA. Here the itineraries are stochastic and one c
ould imagine
that only under extensive ensemble averaging one would obtain tight bounds
on $D$.
In fact we find that using values of $F_1^{(n)}(10^{-8})$ and
$F_1^{(n)}(2\times 10^{-8})$
between 1.002 and 1.01 we can bound $D$ as tightly as $1.6896<D<1.7135$.
Note that
to achieve this accuracy we did not need to go to high values of
$F_1^{(n)}$, but
rather used small values of $\lambda_0$ to reach convergence very early. This
demonstrates again the unexpected fact that the {\em asymptotic} dimension
appears as a renormalization
exponent right after one or a few layers of particles cover the circle, and
very
much before $F_1^{(n)}\sim n^{1/D}$.

At this point we need to address a few questions:

(i) Why classical numerical estimates \cite{83Mea,Mandel} of the fractal
dimension of DLA
converge so slowly?

In standard numerical experiments the radius of gyration of the grown cluster
was plotted in log-log coordinates against the number of particles, with
$D$ estimated from the slope. Examining our fixed point functions $F_1^*$
(see Figs.3-5) we note the slow crossover to linear behaviour, which is not
fully achieved even for extremely high values of $n$. In this respect we
understand
from Eq.(\ref{RvsF}) that a reliable estimate of $D$ from radius of gyration
calculation requires inhuman effort, as was indeed experienced by workers
in the
field \cite{Mandel}. In the present formulation the appearance of the {\em
asymptotic} $D$
as a renormalization exponent already at early stages of the growth allows
a convergent calculation.

(ii) Is a typical DLA cluster self-averaging?

It was demonstrated in \cite{99DHOPSS} that an ensemble of DLA clusters
exhibits
statistics of $F_1^{(n)}$ with standard deviation that shrinks to zero when
$n\to \infty$.
Nevertheless in \cite{00SSB} it was argued that there may remain residual
fluctuations
of the dimension $D$ as extracted from (\ref{eq-scalingrad}). The
numerics presented above is not sufficient to resolve this question, but
additional highly accurate numerics in the limit $\lambda_0\to 0, n\to \infty$
should do.

(iii) Is the problem co-dimension 1?

The multi-scaling properties of the harmonic measure have left the impression
that computing the fractal dimension of DLA will require a simultaneous control
of the host of exponents characterizing the measure. For example the
scaling relation
$D_3=D/2$ (with $D_3$ beeing a generalized dimension in the
Hentschel-Procaccia sense \cite{83HP})
that was derived first by Halsey
\cite{87Hal} strengthened this impression. The approach presented here
indicates
that an appropriate fixed point structure can be
obtained with only one relevant exponent, i.e. $1/D$, and this exponent
appears in the
dynamics much before the measure becomes multiscaling.

\acknowledgments
This work has been supported in part by the
European Commission under the TMR program and the Naftali and Anna
Backenroth-Bronicki Fund for Research in Chaos and Complexity.



\end{document}